\documentclass[%
reprint,
amsmath,amssymb,
aps,
]{revtex4-2}
\usepackage{graphicx}
\usepackage{bm}
\usepackage{multirow}
\usepackage[dvipsnames]{xcolor}

\begin{document}
	
	
	\title{Stochastic gravitational wave background due to core collapse resulting in neutron stars.}
	
	\author{Sourav Roy Chowdhury $ ^{1, 2} $}
	\email{roic@sfedu.ru}

	\author{Maxim Khlopov$ ^{1, 3, 4} $}
	\email{khlopov@apc.in2p3.fr}
	\affiliation{
		$ ^1 $ Research Institute of Physics, Southern Federal University, 344090 Rostov on Don, Russia.\\
		$ ^2 $ Department of Physics, Vidyasagar College, 39, Sankar Ghosh Lane, Kolkata, India.\\
		$ ^3 $ Virtual Institute of Astroparticle Physics, 75018 Paris, France.\\
		$ ^4 $ Center for Cosmoparticle Physics Cosmion, Moscow State Engineering Physics Institute, National
		Research Nuclear University “MEPHI,” 31 Kashirskoe Chaussee, 115409 Moscow, Russia.
	}%

	\date{\today}
	
	\begin{abstract}
		
		The stochastic background of gravitational wave signals arising from the core-collapse supernovae is produced through various complex mechanisms that need detailed and careful investigation. We proposed a simplified multi-peak waveform of the amplitude spectrum. The corresponding energy spectra of our model fit the energy spectra obtained from different numerical simulations of various types of core-collapse events such as non-rotating, slow and fast rotating massive progenitors resulting in neutron stars. The maximum dimensionless energy density $\Omega_{gw}$ corresponding to our model is of $\mathcal{O}(10 {^{-12})} $ around 650 Hz.
		Assuming some degree of uncertainty, we estimated the parameters for the core collapse waveform using BILBY. We studied the detectability of the signal of our model against gravitational-wave detectors like the Einstein Telescope, advanced LIGO and Virgo. Our study indicates that these detectors have to gain more sensitivity to pick up the gravitational wave signals of stochastic background.

	\end{abstract}
	
	\maketitle

	\section{Introduction}

	Gravitational waves (GWs) act as cosmic messengers containing significant information to explain several puzzles in cosmology and astrophysics \cite{thorne, bailes, perkins, abdalla}. GWs can be emitted from a wide range of astrophysical and cosmological sources following the principles of general relativity.
	Various sources exhibit distinct spectral characteristics, reflecting their origins.
	These different characteristics of the relic radiation remain mostly unaltered throughout the propagation. 
	
	In the last few decades, a network of detectors have been constructed to search for compact binary mergers of comparable masses $\textendash$ binary black hole, binary neutron star, or black hole-neutron star \cite{abbott4,abbott5,abbott6}. GW signals detected by the advanced LIGO, Virgo, and KAGRA detectors \cite{abbott3,Abadie, abbott8} offer a venue for tests of general relativity. \cite{gr0,gr1,gr2,gr3}. Such detections of GWs have opened a new window for the \textit{multimessenger astronomy} \cite{mma0,mma1,mma2,mma3,mma5}. Most of the identified mergers are at the redshift, $ z \leq$ 1 \cite{abbott9}, with the network signal-to-noise ratio (SNR) being $\geq12$. However, for all binary black hole analyses, the threshold of the false alarm rate is $<1 ~\textrm{yr}^{-1}$ while that for a binary system with neutron star is $ <0.25~\textrm{yr}^{-1}$ \cite{abbott10}. 
	
	An incoherent superposition of all the unresolved events contributes to the stochastic gravitational wave background (SGWB). It can be astrophysical in origin, sourced by mergers of all binary compact objects \cite{Regimbau2,Regimbau3,Rosado2,Vuk1}, magnetars \cite{Vuk2,Rezzolla1,Lasky,chowdhury}, core-collapses \cite{1,Sandick,Marassi,Crocker,bella} etc. or of cosmological origin, from primordial black holes \cite{Vuk3,Raidal,sah}, early phase instabilities \cite{Owen,Howell}, vacuum fluctuations, \cite{Caldwell,Damour1}, phase transitions \cite{sgb6,sgb7}, cosmic strings \cite{Vuk,sgb8} from the early universe. The most popular approach for detecting a SGWB signal is the \textit{Cross-Correlation} technique developed by Allen et al.  \cite{Allen} which was further modified by Thrane et al. \cite{Thrane1,Thrane2}. There are other search methods, namely \textit{Cross-Cross Correlation Intermediate} (CCI) technique developed by Coyne et al. \cite{Coyne}, \textit{The Bayesian Search} (TBS) technique also developed by Thrane et al. \cite{Thrane3}, etc. A significant improvement in the detection and parameter estimation has been performed in various studies, e.g. Bayesian inference for non-Gaussian SGWB \cite{ref1,ref2}, search for intermittent GW backgrounds \cite{ref3}, incorporation of the off-diagonal correlation of SGWB covariance in the data analysis pipelines \cite{sah}. Despite the availability of numerous search methods, SGWB has not been detected yet \cite{abbott2}. Although, several works \cite{Renzini,Suresh,abbott1} have specified an upper limit on SGWB. Recently, multiple lines of evidence for an isotropic stochastic signal were reported by PPTA \cite{pta}, NANOGrav \cite{nano}, EPTA and InPTA Collaboration \cite{epta} and CPTA \cite{cpta}. The upcoming missions ET, Advanced LIGO, and Virgo have good possibilities to detect SGWB in the Hz - kHz range, while LISA and LGWA are in the sub Hz - Hz range. \cite{caprini,amann,branchesi,babak,harms,ajith}.


	Core-collapse supernovae (CCSNe) are very infrequent events. The estimated rate of these events in the Milky Way is 1.63$\pm$0.46(100 yr)$ ^{-1} $, i.e., the corresponding mean recurrence time is 61$ ^{+24} _{-14} $ yr \cite{milky}. GWs and neutrinos from a supernova (SNe) offer distinct insights into the dynamics of its core-collapse mechanism and set the stage for the initiation of an explosion. Unlike the electromagnetic counterpart that experiences delay and gets developed in a region far from the centre, GWs and neutrinos carry a significant amount of information about the core. In this paper, we solely concentrate on examining SGWB resulting from CCSNe.

	Stars with masses range $\in$[8,25] M$_\odot $ can develop a gravitationally unstable core towards the end of their lives \cite{1st}. 
	During such a process, at different stages, different inhomogeneities are developed, which emit GWs. The core of the stellar object compresses until the density of its inner layer surpasses the nuclear density. At this point, a proto-neutron star (PNS) is developed \cite{Ott1}. As the outer core collapses, it ``bounces" off the interior, producing a shock wave outward. Shock revival is assumed to be the absorption of a fraction of the neutrinos emitted by the PNS \cite{Herant}. Such a process triggers the shock to break down the stellar surface, release the burst, and expose itself as a transient electromagnetic event \cite{Lentz,muller,Janka}. A quasi-periodic signal is expected to be generated while experiencing the convection phase, resulting in a shock. In the post-shock phase, the accretion radius becomes stationary for a period ($\sim$ 300 ms), giving rise to the standing accretion shock instability (SASI) phase \cite{Murphy,Summa,Walk}. This instability, along with neutrinos, acts as a mechanism to transform the gravitational binding energy into the kinetic energy needed for the explosion. Consequently, this leads to a more energetic explosion, being only fueled by neutrinos \cite{1a,Blondin}. A hydrodynamic instability in this phase is expected to be the primary source of low-frequency emission (below 150 Hz, arising in some core-collapse events) \cite{Blondin}, inducing shock oscillations. Another important phase in the later stages of the core bounce is the asymmetric neutrino emission  \cite{Ott,Foglizzo,Takiwaki}. Meanwhile, higher-frequency emission (beyond 500 Hz) is primarily due to oscillations of the PNS.
	
	
	This work emphasizes the study of SGWB due to the stellar core collapse events, which eventually lead to the formation of neutron stars. We proposed a model for the waveform following the principles developed by Buonanno et al. and Sandik et al. \cite{1,Sandick} and estimated the parameters that regulate this waveform to fit the energy spectral distribution of the 3D core collapse simulation results. We took data samples from rapidly rotating, slowly rotating and non rotating massive progenitors with complex hydrodynamics and neutrino transport methods, which are used to study the detection of GW signals.

	The study is structured as follows: 
	In Section \ref{s2}, we proposed a waveform and the associated parameters. In Section \ref{s3}, we discussed the rate estimates of CCSNe. We find the GW energy spectrum of CCSNe of our model in Section \ref{s4}. The detectability of the resulting signal is investigated in Section \ref{s5}, assuming the detectors are LIGO, Virgo, and ET. Finally, we concluded our results in Section \ref{s6}.

	\section{Representation of the CCSNe waveform}\label{s2}

	The detectability of SGWB from a CCSNe requires a reference CCSNe waveform that includes the fundamental characteristics of SNe waves. The quadrupole anisotropy of the system's mass distribution and the strength of sources determine these parameters.
	Following Buonanno et al. \cite{1}, we assumed that every SNe acts as an identical GW source characterized by the Fourier transform.
	\begin{equation}
	\tilde{\texttt{h}}(f) = \int_{-\infty}^{\infty} dt e^{-2 \pi j f t } \texttt{h}(t).\label{char_eqn1}
	\end{equation}
	
	The fact that the energy-momentum tensor does not have a spherical symmetry is reflected through a dimensionless quantity $\texttt{h}(t)$ known as strain amplitude. In most cases, numerical simulations are used to obtain $ \texttt{h}(t) $. Asymmetric neutrino emission \cite{1a} contributes to the strain amplitude through the following equation,
	\begin{equation}
	\texttt{h}(t) = \frac{2G}{c^4 D} \int_{-\infty}^{t-D/c} dt' L_\nu(t') \alpha(t'), \label{char_eqn2}
	\end{equation}
	
	where $ G $ is Newton’s constant, $ c $ is the speed of light, $ D $ is the distance of the SNe to the observer, $ L_\nu(t) $ is the neutrino luminosity, and  $ \alpha(t) $, the asymmetry parameter is defined as the angle-dependent neutrino luminosity, $\leq 1$ \cite{1a}.

	The most notable features emerging from the simulations in the literature are the asymmetric neutrino emission during the explosion and the considerable inhomogeneity produced by the aspherical, large-scale ejecta motion \cite{6,7,8,9,11,12}. This phenomenon contributes to GW production in different stages: (i) while the stellar core bounces against its inner structure, (ii) when the SASI dominates over neutrino-driven convection in the low state (``low state" referring to the post core bounce phase where GW frequency appears in the lower frequency band), and (iii) during the occurrence of PNS oscillations \cite{8,kuroda}. We constructed an empirical model of the waveform to represent a broad spectrum of characteristics in the CCSNe waveform. We employed a multi-peak burst to construct our model of the waveform.  
	In the local frame of a star, the GW amplitude spectrum in the frequency domain ($ f_s $) can be described by the following functional form:		
	\begin{eqnarray}
	f_s|\tilde{\texttt{h}}(f_s)|= \frac{G}{\pi c^4 D} E_{\nu s} <\alpha> & & \Bigg[ 1 + Ae^{{-\big(\frac{f_s-h}{i}}\big)^2} \nonumber \\ & &\hspace{-2.5cm} +  \bigg(1+\frac{f_s}{a}\bigg)^5 e^{-\big(\frac{f_s-p}{q}\big)^2} +Be^{{-\big(\frac{f_s-l}{m}}\big)^2} \Bigg].
	\label{ampli_spec}
	\end{eqnarray}
	
	The energy carried away by the neutrinos during the core collapse and luminosity-weighted averaged neutrino asymmetries are denoted by $ E_{\nu s} $ and $ <\alpha> $ in the source frame, respectively. The first term in parentheses of Eq. (\ref{ampli_spec}) retains the condition in Eq. (\ref{char_eqn2}). The second term represents the low-frequency peak ($ P-I $), while the third  ($ P-II $) and fourth terms ($ P-III $) represent the higher frequency peaks. We limited our study in the frequency band to an upper limit of 4 kHz. We also divided the entire domain into three frequency bins: 10 to 200 Hz, 200 Hz to 2 kHz, and 2 kHz to 4 kHz. For each bin, there are distinct central frequencies ($h,~p \textrm{~and~} l$, respectively) following the CCSNe simulations. $ ~ i,~ q \textrm{~and~} m$ defines the sharpness of the peaks in each bin in Hz, respectively. The free parameters $ A,~a\textrm{~and~} B $ are related to the amplitude in each bin, respectively.
	
	The time-integrated GW energy spectrum from a single core collapse event can, therefore, be determined as
	\begin{eqnarray}
	\frac{dE_{gw}}{df_s} && = \frac{\pi^2 c^3 D^2}{G} (f_s|\tilde{\texttt{h}}(f_s)|)^2 \nonumber \\
	&& = \frac{G}{ c^5 } E_{\nu s}^2 <\alpha>^2 \Bigg[ 1 + Ae^{{-\big(\frac{f_s-h}{i}}\big)^2} \nonumber \\
	&&~~~~~+\bigg(1+\frac{f_s}{a}\bigg)^5 e^{-\big(\frac{f_s-p}{q}\big)^2}  +Be^{{-\big(\frac{f_s-l}{m}}\big)^2} \Bigg]^2.\label{energy_spec}
	\end{eqnarray}

\begin{figure}[htp]
		\includegraphics[scale=0.47]{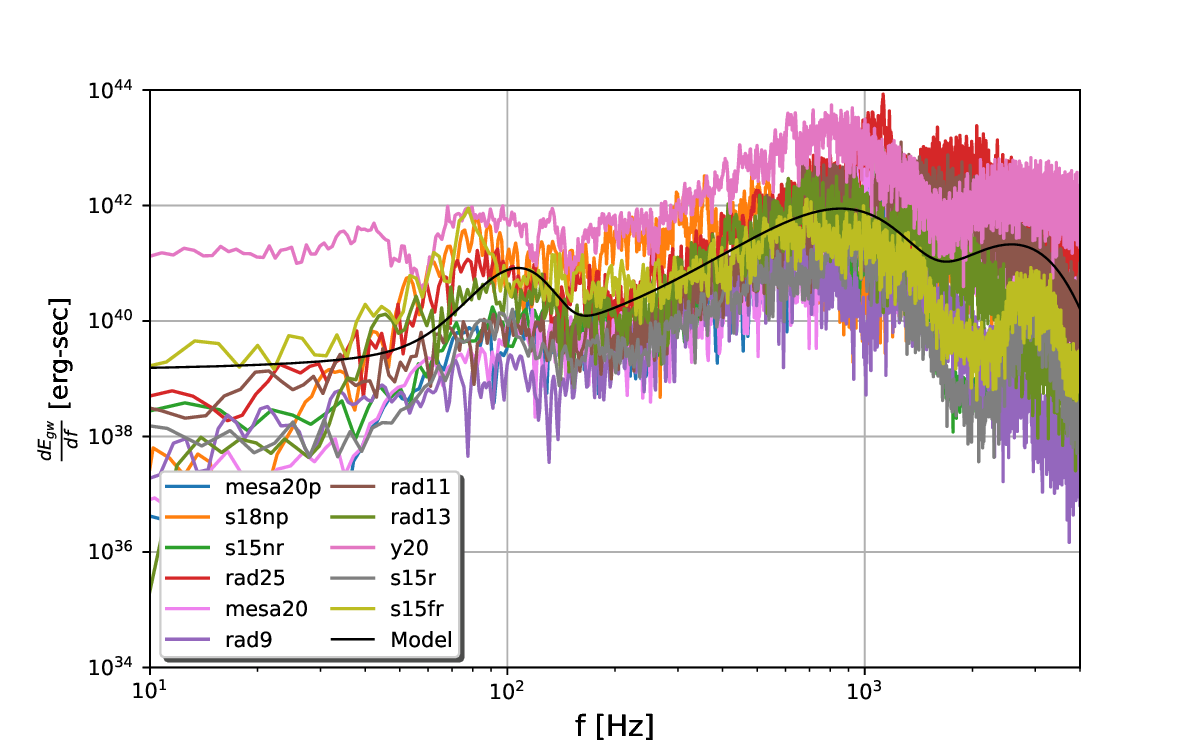}
		\caption{The time-integrated GW energy spectral distribution of the CCSNe assuming multi-peak burst source is shown with respect to frequency. The black solid line curve corresponds to our estimated waveform. Total time-integrated GW spectra $\frac{dE_{gw}}{df}$ for emission from different models [O'Connor and Couch (mesa20, mesa20p) \cite{6}, Radice et al. (rad9, rad11, rad13, rad25) \cite{7}, Powell and M\"{u}ller (y20, s18np) \cite{11,12}, and Andresen et al. (s15nr, s15r, s15fr) \cite{8,9}] are shown with respect to frequency.}\label{energyspect}
	\end{figure}

	The time-integrated GW energy spectra from different simulations and from our model (solid black line) are shown in Fig. (\ref{energyspect}). The parameters used to define our model are provided in Table \ref{t3}.
	
	\section{Star Formation Rate}\label{s3}
	
	Several authors have studied the Star Formation Rate (SFR) extensively, and their estimates converge below redshift, $ z\sim $10 \cite{ins1}. The SFR estimates from high redshift have uncertainties \cite{uncertainity,eff1,eff2} due to the challenges of detecting faint early star-forming galaxies and dust extinction \cite{dust,dust1}.
	
	We considered three different prescriptions of SFR. {\it{First}}, the Madau \& Dickinson star formation rate ({\em{``M-D SFR"}}) \cite{sfr1} for the lower redshift is defined as 
	\begin{equation}
	R_*(z)=0.015\frac{(1+z)^{2.7}}{1+((1+z)/2.9)^{5.6}}.\label{sfr1}
	\end{equation}
	
	{\it{Second}}, the cosmic star formation ({\em{``Cosmic SFR"}}), from the ``dark ages'' at redshift, $ z $ = 15 to the present, following Springel \& Hernquist \cite{sfr2} as follows,
	\begin{equation}
	R_*(z)=\nu \frac{ \alpha \exp(\beta (z - z_m))}{\alpha - \beta + \beta \exp(\alpha (z - z_m))}.\label{sfr2}
	\end{equation}
	
	The optimum redshift for the Cosmic SFR is described by ${z_m}$, while $ \nu $ represents the amplitude (astration rate), in $ M_\odot yr^{-1} Mpc^{-3} $. The SFR slope for the low and the high redshifts are denoted by $\beta$ and $(\beta - \alpha)$, respectively. The parameters chosen for the Eq. \eqref{sfr2} are $ \nu = 0.15~M_\odot yr^{-1} Mpc^{-3} , z_m = 5.4, \alpha = 0.933 \textrm{~and~} \beta = 0.66 $ .
	
	\begin{figure}[htp]
		\includegraphics[scale=0.47]{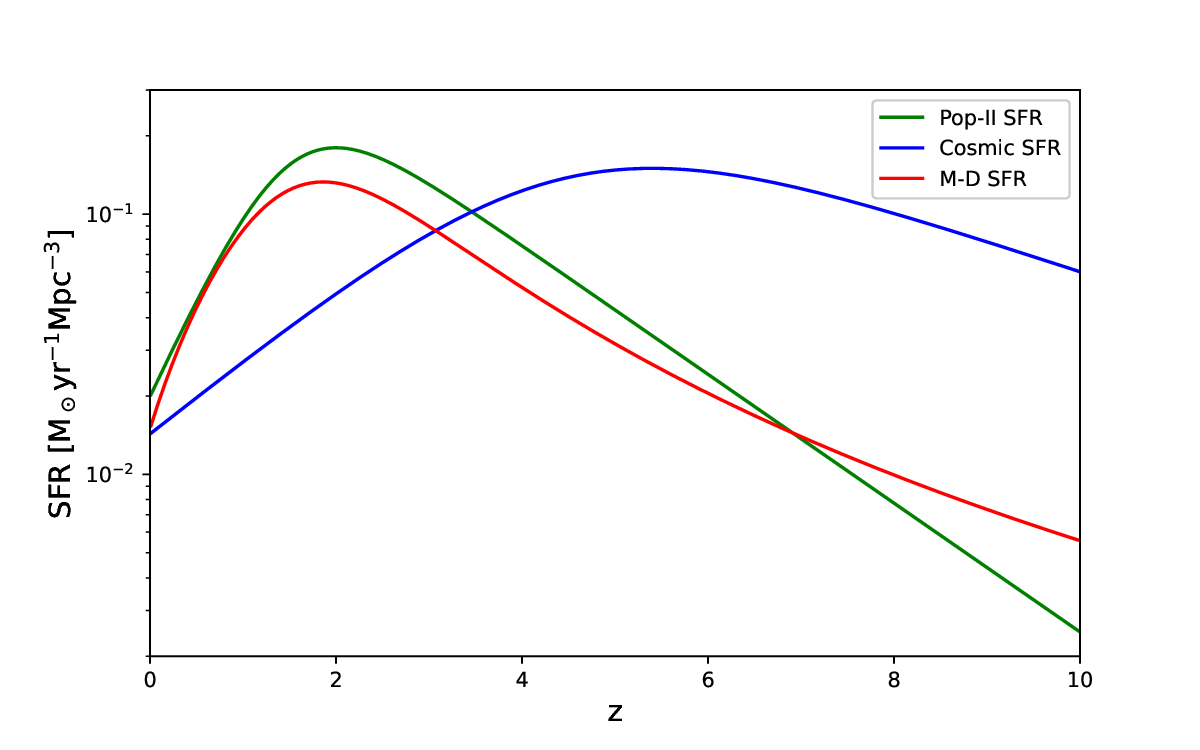}
		\caption{SFR as a function of redshift for different models namely {\it{M-D SFR}}, {\it{Pop-II SFR}} and {\it{Cosmic SFR}} are explored here. The {\it{M-D SFR}} and the {\it{Pop-II SFR}} models result higher SFR for lower redshifts unlike {\it{Cosmic SFR}} that results higher SFR at high redshifts}\label{sfr}
	\end{figure}

	{\it Third,} a standard mode of the population II stars formation  ({\em{``Pop-II SFR"}}) between 0.1 M$_\odot$ and 100 M$_\odot$, following Springel \& Hernquist \cite{sfr3}. Cosmic SFR and Pop-II SFR have the same functional form with different parametric values. The parameter set considered for the Pop-II SFR are $ \nu = 0.18~M_\odot yr^{-1} Mpc^{-3} , z_m = 2.0, \alpha = 2.37 \textrm{~and~} \beta = 1.8 $.

	The variation of SFR with redshift $ z $ from different prescriptions is shown in Fig. \ref{sfr}. The rate of core-collapse events per unit {\em{comoving volume}}, $R_v (z)$ that contributes to GW energy density is directly related to the SFR (i.e., there is no time delay) as follows,
	\begin{equation}
	R_v(z) = \lambda_{cc}  R_*(z). \label{rate_sne}
	\end{equation} 
	
	Here, $ \lambda_{cc} $ is the mass fraction of stars \cite{10} that experience core collapse, in units of $ M^{-1}_\odot$. Stars in the mass range [8,25] M$_\odot $ undergo core-collapse processes to become neutron stars. For an initial mass function, we consider Salpeter form $ \phi(m) \propto m^{-2.35} $.  
	
	Therefore, the estimated mass fraction that undergoes core collapse is
	\begin{eqnarray}
	\lambda_{cc} = \frac{\int_{8M_\odot}^{25M_\odot}\phi(m) dm}{\int_{0.1M_\odot}^{\infty} m \phi(m) dm} \approx 0.00549 M^{-1}_\odot.\label{lambda_cc}
	\end{eqnarray}
	
	It is assumed to be the same for all redshifts. This free parameter represents a scaling parameter associated with the CCSNe rate. 
	
	\section{SGWB from astrophysical sources}\label{s4}

	The normalized GW energy density can be represented by the dimensionless quantity $\Omega_{gw}(f)$. Following \cite{Allen}, $\Omega_{gw}(f)$ can be written as
	\begin{equation}
	\Omega_{gw}(f) = \frac{1}{\rho_c} \frac{d \rho_{gw}}{d \ln f}.
	\end{equation}
	
	Here, $\rho_{gw}$ is the energy density within a frequency band $f \rightarrow f +df$ in SGWB. $\rho_c$ is the critical energy density at the present time, which is related to the Hubble constant $H_0$ through the equation 
	\begin{equation}
	\rho_c = \frac{3 H_0^2 c^2}{8 \pi G}.\label{crit_edensity}
	\end{equation}
	
	Following \cite{Regimbau2, Vuk1}, for CCSNe background, the total GW energy density spectrum from all sources measured in the source frame is related to $dE_{gw}/df_{s}$ in Eq. \eqref{energy_spec} and $ R_v(z) $ in Eq. \eqref{rate_sne} as	
	\begin{equation}
	\Omega_{gw}(f) = \frac{f}{\rho_c H_0} \int_{0}^{\infty}  \frac{R_v(z) dz}{(1+z) E(\Omega, z)} \frac{d E_{gw}}{df_s}.\label{spec}
	\end{equation}
	
	$ f_s $ is the frequency of the wave measured in the source rest frame. It is related to the frequency $f$ measured in the detector frame as $ f_s = f(1+z) $. The function $E(\Omega, z)$ is defined as,
	\begin{equation}
	E(\Omega, z) = \sqrt{\Omega_r (1+z)^4 +\Omega_m (1+z)^3+\Omega_\Lambda}, 
	\end{equation}
	
	where, $\Omega_r = 9.094 \times 10^{-5}$, $\Omega_m = 0.315 \pm 0.007$, $\Omega_\Lambda = 0.685 \pm 0.007$ and $H_0 = 67.66 \pm 0.42$ km/s/Mpc are the standard $\Lambda$CDM parameters \cite{planck}. They represent the density parameters for radiation, pressureless matter, cosmological constant and the value of the Hubble constant, respectively.
	
	The explicit form of the GW energy density for CCSNe measured in the detector frame is as follows,
	\begin{eqnarray}
	\Omega_{gw}(f) = &&\frac{8 \pi G f \xi}{3 c^2 H_0^3} \int_{0}^{\infty}  \frac{R_*(z) dz}{ (1+z)^3 E(\Omega, z)} 
	\Bigg[ 1 + Ae^{{-\big(\frac{f(1+z)-h}{i}}\big)^2}\nonumber \\
	&& \hspace{1.85cm}+\bigg(1+\frac{f(1+z)}{a}\bigg)^5 e^{-\big(\frac{f(1+z)-p}{q}\big)^2} \nonumber \\
	&& \hspace{3.7cm} +Be^{{-\big(\frac{f(1+z)-l}{m}}\big)^2} \Bigg]^2, \label{final_omega}
	\end{eqnarray}
	
	where, 
	$\xi = \frac{G}{c^5} E_{\nu s}^2 <\alpha>^2 \lambda_{cc} $, measured in $ m^2/s $.  This parameter is simply a scaling factor corresponding to the total energy emitted in GWs. Overall amplitude of the waveform depends on $ \xi $, a higher $ \xi $ typically accompanies a higher amplitude.
	
	
	\begin{figure}[htp!]
		\includegraphics[scale=0.47]{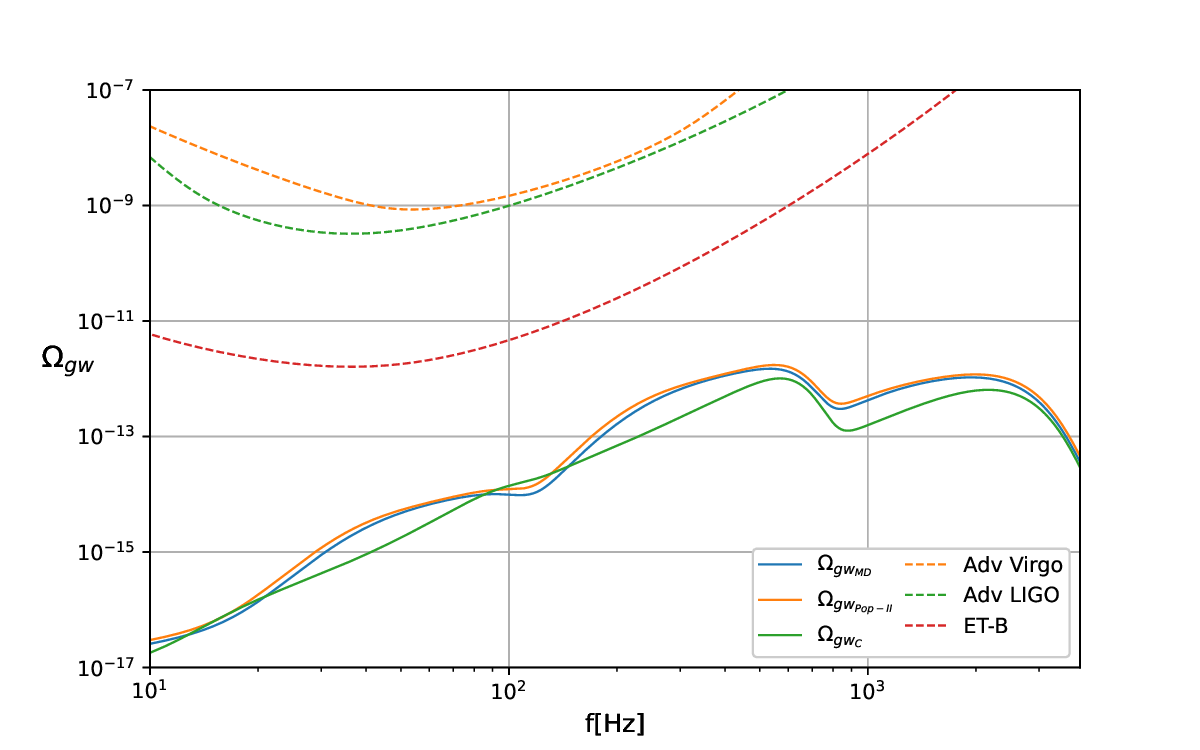}
		\caption{The variation of the energy density spectra $\Omega_{gw}(f)$ is shown for different $ R_*(z) $, with frequency. $ \Omega_{gw_{{M-D}}} $, $ \Omega_{gw_{Pop-II}} $ and $ \Omega_{gw_{C}} $ stands for energy density spectra for {\it{M-D SFR}}, {\it{Pop-II SFR}} and {\it{Cosmic SFR}}, respectively. Moreover, considering a year of exposure and co-located detectors, the SNR = 1 for advanced LIGO, Virgo and ET are presented. }\label{Omega}
	\end{figure}

	The energy density spectra obtained from our model are defined in Eq. \eqref{final_omega}. We plotted $\Omega_{gw}(f)$ for different SFRs as solid curves in Fig. \ref{Omega}. We noted that the overall contribution of a distant system (higher redshift) and a less distant (lower redshift) system is about the same. The dashed curves represent the stochastic background calculated for the network of the advanced Virgo, LIGO and ET(ET-B), respectively, in the figure. We assumed that the detectors are co-located, with one year of exposure and SNR = 1. Detailed analysis of SNR for our model is provided in Sec. \ref{s51}.

	\section{DETECTABILITY WITH ADVANCED DETECTORS}\label{s5}
	
	Search for a stochastic background can pick up significant errors from the intrinsic background noise of the detectors. In comparison to random detector ``noise", SNR quantifies how strong the stochastic background is. We consider results from three advanced detectors to assess SGWB detectability for CCSNe. The continuity of a signal can be justified by the duty cycle.
	
	\subsection{SNR}\label{s51}
	Cross-correlating measurements from different detectors is an ideal approach to separate SGWB signals from detector background noise. Following Allen and Romano \cite{Allen}, the optimal SNR for SGWB, for a given integration time $ T $ can be found using,
	\begin{equation}
	SNR = \frac{3H_o^2}{10 \pi^2} \sqrt{2T} \Bigg[\int_{0}^{\infty} \frac{\gamma^2(f)\Omega_{gw}^2(f)}{f^6 P_1(f) P_2(f)}df\Bigg]^{1/2},\label{snr}
	\end{equation}
	
	where $\gamma(f)$ is called a normalized overlap reduction function that characterizes the loss of sensitivity caused by separation and relative orientation of the detectors. $ P_1(f) $ and $ P_2(f) $ are the power spectral noise densities of two detectors.

	\begin{table}[htp]
		\caption{SNR of the model for three different SFRs. The table gives values for advanced LIGO, Virgo and ET-B.}\label{t1}
		\begin{ruledtabular}
			\begin{tabular}{ c  c  c  c }
				\textbf{Detector~~}   & {\textbf{~~Pop-II~~}}  & {\textbf{~~~~M-D~~~~}} & {\textbf{~~Cosmic~~}}\\ 
				\hline
				
				Adv LIGO $\times$ 10$ ^{-3}  $ & 3.7166  & 3.1867  & 2.0575 \\ 
				
				Adv Virgo $\times$ 10$ ^{-3}  $ & 1.2122  & 1.3179  & 0.6972 \\ 
				
				ET-B $\times$ 10$ ^{-1}  $ & 1.6764  & 1.4807 & 0.6153 \\  			
			\end{tabular}
		\end{ruledtabular}
	\end{table} 
	
	In this study, the observation period is considered to be one year. For our study, $ \gamma(f) $ is fixed to 1 and (-3/8) for the co-located detector pair (for advanced LIGO and Virgo) and ET-B, respectively \cite{orf,orf1}. The values of SNR for our model for different SFRs are shown in detail in Table-\ref{t1}.	
	
	\subsection{Duty Cycle}
	The duty cycle (DC) is a useful quantity for the characterization of an SGWB of astrophysical origin. It is the ratio between the typical duration of an individual cycle and the average time interval within two consecutive events. Following Regimbau and Mandic \cite{10}, the DC is
	\begin{equation}
	DC =  \frac{4 \pi \lambda_{cc} c }{H_0} \int_{0}^{z_{max}} \frac{\tau R_*(z)  D^2(z)}{E(\Omega,z)} dz. \label{dc}
	\end{equation}
	
	$ \tau $ is the average duration of individual bursts starting from the time of emission, which we fixed at 1 ms.
	
	A signal with unity or higher DC for a CCSNe is regarded as continuous. Non-continuous signals can be classified into short and popcorn types using the Regimbau \& de Freitas Pacheco convention \cite{Regimbau}. The total duty cycles for {\it{Cosmic SFR}}, {\it{Pop-II SFR}} and {\it{M-D SFR}} are 2.81611, 0.84074 and 0.68057 respectively.

	\section{Summary and Discussion}\label{s6}
	
	In this work, we have investigated the stochastic GWs that is produced by stellar core collapse events leading to neutron star formation throughout the universe. Since the understanding of the waveform of the GW emitted during a single core collapse event is limited, we utilized an empirical functional form to model the waveform. We constructed an empirical waveform of multi-peak CCSNe. The amplitude and the quality factor at each central frequency are estimated as the weighted parameters using state-of-the-art simulation models. A low-frequency peak is predicted due to acoustic waves produced by prompt convection (post-core bounce state) at around 95 Hz, the most sensitive frequency range for stochastic search. The successive high-frequency peaks in the emission spectrum are most likely due to the large-scale, spherical ejecta motion and asymmetric neutrino emission. Furthermore, the simulations show several other higher-frequency peaks which are beyond the detectors' range. 
	
	
	
	


	
	The core-collapse contribution to $\Omega_{gw}$ at the lower bound ($\approx$10 Hz) of the inspected frequency spectrum is $\mathcal{O}(10 ^{-16})$. In this low-frequency range, the contribution from core-collapse GW background is challenging to separate from cosmological GW backgrounds \cite{sgb7,sgb8,Maggiore,cgwb}. In comparison, the maximum contribution of core-collapse GW background to $\Omega_{gw}$ is $\mathcal{O}(10 ^{-12})$ at $\approx$ 650 Hz, have a high possibility of detection. We noted that the overall contributions of several SFRs are nearly the same. However, it is found that  {\it{M-D SFR}} and {\it{Pop-II SFR}} support non-continuous signal, whereas {\it{Cosmic SFR}} supports the continuous signal, attributing to its effectively large redshift. We compared the resulting spectra from our model to the anticipated sensitivities of the advanced detectors. The sensitivity predicted by our model is $\mathcal{O}(10^{-1})$ and $\mathcal{O}(10^{-3})$ times the considered sensitivity of ET and the detector pair (advanced LIGO and Virgo), respectively. Our model is well-constrained in the low frequency region (1-90 Hz). However,	more numerical core-collapse simulations are required to constrain a broader frequency range (10$ ^{-4} $-100 Hz).
	


	CCSNe are inevitably accompanied by neutrino radiation. We can approximate the expected flux of these neutrinos from the estimated energy of neutrino radiation from individual SNe \cite{n1}. It contains information of the dynamical processes and properties of the progenitor \cite{Halzen}. Spherically symmetric collapse can also lead to neutrino radiation. Correlated detections of neutrinos and GWs lead to the signature of asymmetric CCSNe and a potential reduction of sky localizations. Developing specific methods of analysis, including machine learning, may be a potential treatment to detect GWs \cite{ml1}. Algorithms can be trained to isolate signals from background noise \cite{ml2,ml4}, to improve accuracy and robustness \cite{ml}, to refine input parameters and physical models used \cite{ml3}, etc. It provides a possibility to study \textit{multimessenger} astronomy \cite{mma0,mma1,mma2,n2,n4}.

	

	\section*{ACKNOWLEDGMENTS}
	
	The authors thank the referee for the constructive report and feedback. SRC would like to thank Vuk Mandic and Haakon Andresen for their valuable discussions during the early stages of this study and Bella Finkel for providing the data. The authors warmly thank Paul Lasky for various discussions and communications on the topic. 
	The Southern Federal University supported the work of SRC (SFedU). The research by MK was carried out at Southern Federal University with financial support from the Ministry of Science and Higher Education of the Russian Federation (State contract GZ0110/23-10-IF).

	\appendix
	
	\section{Estimation using BILBY}
	
\begin{figure*}[htp]
		\includegraphics[scale=0.55]{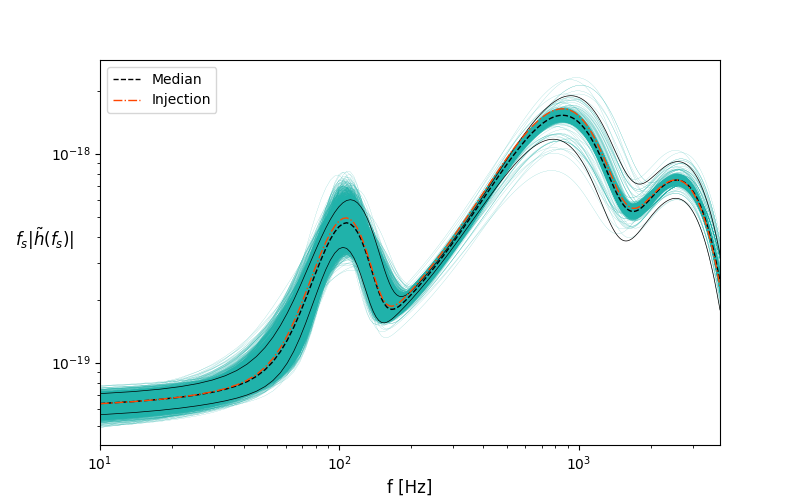}
		\caption{The band represents the \textit{multi-peak} waveform reconstructions derived from the posterior samples using a Gaussian likelihood function. The red dash-dotted line indicates the injected waveform. The solid black lines denote the 5-95\% confidence interval and the dashed black line denotes the median.}
		\label{waveform}
	\end{figure*}

	For a more detailed analysis of the parameters of the waveforms, we performed Bayesian parameter estimation. Using BILBY \cite{bilby}, we replicated waveform of CCSNe and estimated the parameters. The proposed multi-peak waveform in Eq. \ref{ampli_spec} is a function of parameters ($ A, h, i, a, p, q, B, l, m, \xi $). Eq. \ref{ampli_spec} is implemented into BILBY as a waveform model. We used a standard Gaussian noise probability for the waveform. The priores implemented for the CCSNe system are provided in Table. (\ref{t3}). For the sake of simplicity, we assumed uniform priors for each parameter. In this example, we used the \texttt{GaussianLikelihood} and \texttt{Dynesty} sampler. The posterior predictive waveform with the injected (red) waveform is shown in Fig. \ref{waveform}. The shaded (teal) region represents a superposition of many reconstructed waveforms from the posterior samples.
	
\begin{figure*}[htp]
		\includegraphics[scale=0.33]{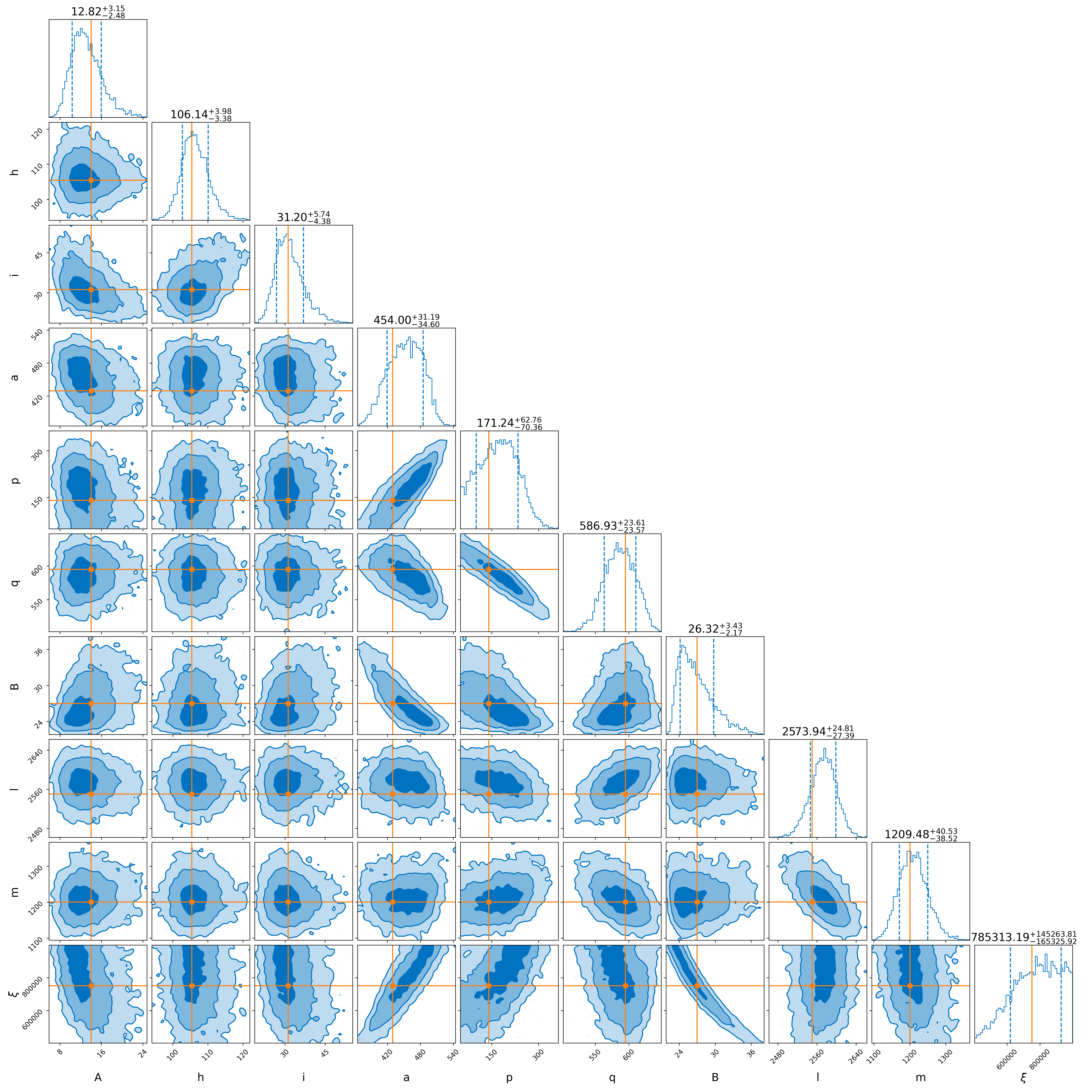}
		\caption{Corner plot illustrates the posterior parameter distribution for a CCSNe waveform. Along the diagonal, there are one-dimensional marginalized posteriors for each fitting parameter. In the off-diagonal plots, two-dimensional maps showcase the posterior distributions marginalized over all alternative parameters. Orange vertical and horizontal lines indicate the injected parameter values of the simulated waveform.}\label{SN}
	\end{figure*}

	Fig. \ref{SN} displays the recovered posterior distributions (the marginals and all pairwise correlations) of our parameters to a CCSNe waveform. Orange lines in the plot represent the injected samples. The measurement uncertainties are assumed to be Gaussian in nature. The parametric values used in this study are within  1$\sigma$ uncertainty with no flux.



	\begin{table}[htp]
		\caption{Estimation of the parameters of the waveform.}\label{t3}
		\begin{ruledtabular}
			\begin{tabular}{ c  c  c  c}
				\textbf{Variable~}   & {\textbf{~Injected values~}}  &  {\textbf{~Bounds~}} & {\textbf{~Posterior~}} \\ 
				\hline
				
				$ A $ & 14  & [0.1, 35]& $ 12.82^{+3.15}_{-2.48} $ \\ 
				
				$ h [\textrm{Hz}]$ & 105  & [60, 140] & $ 106.14^{+3.98}_{-3.38} $ \\ 
				
				$ i [\textrm{Hz}]$ & 31  & [1, 50] & $ 31.20^{+5.74}_{-4.38} $ \\  
				
				$ a [\textrm{Hz}]$ & 430  & [100, 800] & $ 454.00^{+31.19}_{-34.60} $ \\  
				
				$ p [\textrm{Hz}]$ & 140  & [50, 500] & $ 171.24^{+62.76}_{-70.36} $ \\  
				
				$ q [\textrm{Hz}]$ & 595  & [100, 900] & $ 586.93^{+23.61}_{-23.57} $ \\  
				
				$ B $ & 27  & [1, 50] & $ 26.32^{+3.43}_{-2.17} $ \\  
				
				$ l [\textrm{Hz}]$ & 2550  & [1000, 3500] & $ 2573.94^{+24.81}_{-37.39} $ \\  
				
				$ m [\textrm{Hz}]$ & 1200  & [600, 1800] & $ 1209.48^{+40.53}_{-38.52} $ \\  
				
				$ \xi$ [\textrm{m$^2 $/s} ] & 7.5$\times10^5$  & [4, 10]$\times10^5 $  & $ 785313^{+145263.81}_{-165325.92} $ \\  			
			\end{tabular}
		\end{ruledtabular}\\
		\vspace{0.2cm}
		Note: The given posterior values of each parameter are the median, 5\%, and 95\% quantiles.
	\end{table}

\end{document}